\documentclass[twocolumn,prb,aps,superscriptaddress,showpacs]{revtex4}
\usepackage{graphicx}
\begin{document}
\title {Interaction of a surface acoustic wave with a two-dimensional electron gas}
\author{Shi-Jie Yang}
\author{Hu Zhao}
\affiliation{Department of Physics, Beijing Normal University,
Beijing 100875, China}
\author{Yue Yu}
\affiliation{Institute of Theoretical Physics, Chinese Academy of
Sciences, P.O. Box 2735, Beijing 100080, China}

\begin{abstract}
When a surface acoustic wave propagates on the surface of a GaAs
semiconductor, coupling between electrons in the two-dimensional
electron gas beneath the interface and the elastic host crystal
through piezoelectric interaction will attenuate the SAW. The
coupling coefficient is calculated for the SAW propagating along
an arbitrary direction. It is found that the coupling strength is
largely dependent on the propagating direction. When the SAW
propagates along the [011] direction, the coupling becomes quite
weak.

\end{abstract}
 \maketitle

\section{introduction}
Experiments which include coupling of surface acoustic waves
(SAW's) with electrons in a two dimensional electron gas (2DEG)
have intrigued great interest in these years. SAW experiments that
have been carried out in the fractional quantum Hall regime near a
filling factor $\nu=1/2$, strongly support the composite fermion
approach to the compressible state at $\nu=1/2$\cite{Halperin}. In
quantum Hall systems the formation of modulated electron
structures can occur at half-filled high Landau levels. These new
phases was predicted theoretically and confirmed experimentally by
observation of a strong anisotropy of the conductivity. The
piezoelectric coupling is considered as the origin of the
orientation of stripes or electron crystal\cite{Sher,Fil}.

On the other hand, SAW has also been demonstrated as another
method of creating electron wires in 2DEG\cite{Govorov,Alsina}.
This method involves an intense SAW in a hybrid structure
containing a semiconductor quantum-well and a strongly
piezoelectric coupling host crystal. In the experiments a
homogeneous 2DEG in the hybrid structure turns into moving
electron wires due to a very strong piezoelectric potential
induced by a SAW. This effect has been demonstrated in the above
mentioned experiments for the classical regime of
acoustic-electric interaction at room-temperature. For low
temperatures and a sufficiently strong piezoelectric potential of
a SAW, the electron spectrum of wires can be quantized in the
direction of the SAW momentum\cite{Aharony,Tal}.

It is well-known that the propagation of acoustic wave on the
surface of piezoelectric crystals can be effected by nearby
conductors\cite{Ing}. In a heterostructure where a 2-dimensional
electron gas forms beneath the interface, the interaction of the
electron gas with the SAW's will the attenuate the propagation
with a velocity shift $\delta v$ and the attenuation coefficient
$\kappa$, which satisfy the relation\cite{Efros}
\begin{equation}
\frac{\delta
v}{v}-\frac{i\kappa}{q}=\frac{\alpha^2/2}{1+i\sigma_{xx}(q,\omega)/\sigma_m}
\label{no1}
\end{equation}
where $\sigma_{xx}(q,\omega)$ is the longitudinal conductivity of
the adjoining medium at wave vector $q$ and frequency $\omega=v
q$. The coefficients and $\alpha^2/2$ depend on material
parameters which will be discussed in this work.

Unlike conductivity measurements in which a current is driven
through the 2DEG and the voltage is measured, where only the
conductivity at zero wave vector can be probed, SAW measurements
allow one to probe the finite wave vector
conductivity\cite{Willett}. In earlier SAW experiments, the
wavelength of the SAW was much larger than the distance $d$ of the
2DEG from the surface. In this case, the depth can be neglected,
and the coefficients $\sigma_m$ and $\alpha^2/2$ can be assumed to
be constant. When the wavelength is comparable to the distance
$d$, one should be ware of the wave-vector dependence of the
coefficients\cite{Willett,Willett2}. Roughly one might expect that
the coupling $\alpha^2/2$ should decay approximately as
$e^{-2qd}$.

In this work, we use a model which takes into account the
anisotropy of the elastic constants of the crystalline matrix. The
physical properties of the GaAs system are described by three
elastic constants $c_{11},c_{12}$, and $c_{44}$, one piezoelectric
modulus $e_{14}$, and the dielectric constant $\epsilon $. We
study the propagation of SAW along an arbitrary direction with and
without piezoelectric coupling where the 2DEG is located a
distance $d$ below the surface, as shown in Fig.1. In Sec. II we
describe the transport of the SAW and in Sec. III introduce the
energy shift by piezoelectric coupling. The coupling coefficient
of the SAW with 2DEG is calculated in Sec. IV. A brief discussion
is given in Sec. V.

\section{Transport of the SAW}
Consider an infinite piezoelectric medium. We study the elastic
wave movement by defining a displacement vector $u_k$. The elastic
stress tensor satisfies $T_{ij}=c_{ijkl}u_{kl}-e_{kij}E_k$, where
$u_{kl}=\frac{1}{2} (\partial_k u_l+\partial_l u_k)$ and $\bf{E}$
is electric field vector\cite{Landau}. $c_{ijkl}$ is the elastic
modulus which in cubic symmetry crystals have only three
independent parameters $c_{11}$, $c_{12}$, and $c_{44}$. For GaAs
at low temperatures, these elastic constants are measured as
$12.26\times 10^{10}$, $5.71\times 10^{10}$, and $6.00\times
10^{10}$ N/m$^2$, respectively. $e_{kij}$ is the piezoelectric
stress tensor which has only one nonzero component $e_{14}$ in
GaAs and AlAs crystals. It has an accepted value of approximately
$0.15$ C/m$^2$ for GaAs. The electric displacement vector
$D_j=e_{jkl}u_{kl}+\epsilon E_j$ by taking account of the
piezoelectric coupling. Here $\epsilon$ is the dielectric constant
of the medium which is assumed to be isotropic. The elastic wave
equation is given by
\begin{eqnarray}
c_{ijkl}\partial_l\partial_j u_k +e_{kij}\partial_j\partial_k \phi
&=&\rho \partial_t^2 u_i  \nonumber\\
e_{jkl}\partial_l\partial_j u_k-\epsilon\nabla^2\phi &=&0,
\label{motion}
\end{eqnarray}
where $\rho$ is the mass density. For GaAs crystal, $\rho=5300$
kg/m$^3$.

We first consider a SAW propagates on a free surface by
disregarding of the piezoelectric coupling, i.e., the crystal is
non-piezoelectric. The boundary condition in the $\hat{z}$
direction is
\begin{equation}
T_{\hat{z}k}=c_{\hat{z}klm}\partial_l u_m|_{z=0}=0.
\label{boundary}
\end{equation}

A surface acoustic wave is propagating along an arbitrary
direction with an azimuth angle $\theta$ to the $\hat{x}$ axis.
The plane wave solution has the following form
\begin{equation}
u_i=u_{0i} exp[-i\omega t+i{\bf q}\cdot {\bf r}+i q_z z],
\label{trial}
\end{equation}
where ${\bf q}=(q_x,q_y)$ is the wave vector in the $X-Y $ plane
and $q_z$ is vector in the $\hat{z}$ direction. Substitute formula
(\ref{trial}) into the first line of Eqs.(\ref{motion}) by
omitting the piezoelectric coupling term, one obtains
\begin{equation}
[c_{ijkl}q_lq_j-\rho \omega^2\delta_{ik}]u_{0k}=0. \label{no2}
\end{equation}
For surface acoustic waves, $q_z$ should be a complex quantity
which implies the wave decay exponentially into the bulk. The
eigenvalues of the wave vectors $q_z^{(n)}$ and the respective
eigen solutions $u_{0i}^{(n)}$ ($n=1,2,3$) are determined by
Eq.(\ref{no2}) . To satisfy the boundary condition
(\ref{boundary}), a linear combination of the eigen solutions
should be taken
\begin{equation}
u_i=\sum_{n=1}^3 C(n) u_{0i}^{(n)} \exp^{[-i\omega t+i{\bf q}\cdot
{\bf r}+iq_z^{(n)}z]}. \label{velocity}
\end{equation}
From the above equation, the velocity of the SAW and relevant
coefficients of $C(n)$ are computed. Figure 2 displays a
non-monotonous relation of $v_s$ with the propagation direction in
a GaAs crystal. The velocity $v_s$ reaches a peak when the SAW
transports along the direction of $\theta\sim 25^0$ while at
$\theta=45^0$ it falls to a minimum.

\section{interaction-induced energy shift}
Due to the piezoelectric coupling, an external scalar potential
$\phi^{ext}=A e_{14}F(qd)/\epsilon$ is induced in the 2DEG, where
$A$ is the amplitude of the SAW and F is a dimensionless function
of $qd$ that represents the fact that the SAW decays into the bulk
of the crystal. The induced energy density per unit area due to
this external potential is given by\cite{Simon}
\begin{equation}
\delta U=\frac{\epsilon_{eff}q}{4\pi
[1-i\sigma_m/\sigma_{xx}(q,\omega)]}|\phi^{ext}|^2,
\end{equation}
where $\epsilon_{eff}$ is the effective background dielectric
constant which is wave-vector-dependent with the form
\begin{equation}
\epsilon_{eff}/\epsilon=\frac{1}{2}[\frac{(\epsilon+1)\exp
(qd)}{\epsilon \cosh(qd)+\sinh(qd)}].
\end{equation}
We want to measure this energy shift with respect to the shift for
$\sigma_{xx}\rightarrow \infty$.
\begin{equation}
\Delta U\equiv \delta U-\delta U(\sigma_{xx}\rightarrow \infty).
\end{equation}

It is found below that the surface acoustic wave has an energy
density proportional to $A^2q^2$. Furthermore, the wave decays
exponentially into the bulk with a decay constant proportional to
$q$. Thus, when integrated in the $\hat{z}$ direction, the energy
$U$ per unit surface area is given by
\begin{equation}
U=qA^2H,
\end{equation}
where $H$ is a factor that depends on material parameters and is
determined by calculating the energy density induced by SAW.

Combined above results, the fractional energy shift is then given
by
\begin{equation}
\frac{\Delta
U}{U}=\frac{-\alpha^2/2}{1+i\sigma_{xx}(q,\omega)/\sigma_m},
\end{equation}
where the coefficient $\alpha^2/2$ in Eq.(\ref{no1}) is derived as
\begin{equation}
\alpha^2/2=\frac{\epsilon_{eff}}{\epsilon}\frac{e^2_{14}}{\epsilon_0
\epsilon H}|F(qd)|^2. \label{alpha}
\end{equation}

\section{piezoelectric coupling with 2DEG}
Since the piezoelectric coupling $e_{14}$ is small, it is seen
from the second equation of Eqs.(\ref{motion}) that $\phi$ is of
order $e_{14}$ smaller than $u$. Thus the first equation will be
solved by the non-piezoelectric solution discussed above with
corrections only at order $e_{14}^2$. The mechanical boundary
conditions in the piezoelectric case are
\begin{equation}
c_{i\hat{z}kl} \partial_l u_k +e_{ik\hat{z}}\partial_k\phi=0.
\end{equation}
The electrical boundary condition for the second equation of
Eqs.(\ref{motion}) is determined by setting the potential to be
continuous at $z=0$,
\begin{equation}
e_{\hat{z}kl}\partial_l u_k-\epsilon\partial_z\phi
|_{z=0^-}=-\partial_z\phi |_{z=0^+}.
\end{equation}

The piezoelectric case for $\phi$ is solved by make use of the
non-piezoelectric solution for displacement $u$ in Section II. The
supposed form of solution is
\begin{equation}
\phi=\frac{e_{14}}{\epsilon}e^{i({\bf q}\cdot{\bf r}-\omega
t)}[A(1)e^{iq_z^{(1)}z} +A(2)e^{iq_z^{(2)}z} +A(3)e^{iq_z^{(3)}z}
+A(4)e^{q z}]. \label{phi}
\end{equation}
Substitute the above solution and formula (\ref{velocity}) into
Eq.(\ref{motion}), one gets
\begin{eqnarray}
A(n)=&&\frac{1}{q^2+(q_z^{(n)})^2}[C(n)u_{01}^{(n)}q_yq_z^{(n)}\\\nonumber
&&+C(n)u_{02}^{(n)}q_xq_z^{(n)}+C(n)u_{03}^{(n)}q_xq_y]\\\nonumber
&&(n=1,2,3)
\end{eqnarray}
and
\begin{eqnarray}
A(4)=&&\frac{1}{q(\epsilon+1)} \sum_{n=1}^3 [i\epsilon
A(n)q_z^{(n)}-q A(n)\\\nonumber
&&-\frac{i\epsilon}{2}(q_xC(n)u_{02}^{(n)}+q_yC(n)u_{01}^{(n)})],
\end{eqnarray}
where $A(1)=-A(2)^*$ and $A(3)$, $A(4)$ are pure imaginary. The
explicit form of the potential solution is then
\begin{eqnarray}
\phi=&& \frac{iAe_{14}}{\epsilon} [2 Im (A(1)e^{iq_z^{(1)}z})
\\\nonumber
&&+Im A(3)e^{iq_z^{(3)}z}+Im A(4)e^{qz}]e^{i{\bf q}\cdot{\bf
r}-i\omega t}.
\end{eqnarray}
Thus we obtain the dimensionless function $F$ in formula
(\ref{alpha}) as
\begin{equation}
F(q)=2|A(1)|e^{-\beta d}\sin (\xi-\alpha d)+Im (A(3))e^{-\gamma
d}+Im(A(4))e^{-qd},
\end{equation}
where $\alpha$, $\beta$ and $\xi$ are the real part, imaginary
part and phase of $A(1)$, respectively.

Figure 3 shows the dependence of the coupling coefficient
$\alpha^2/2$ on the orientation angle that the SAW propagates in a
GaAs crystal. They are exponentially decay into the bulk of the
sample, as we had expected. However, For different transport
angles, the coefficients differ largely. They also exhibit a
non-monotonous relation with the angle. For example, near the
region of $\theta\sim 25^0$, the coefficient reach the maximum
while when the SAW transport along the direction of $\theta=45^0$,
the coupling between the 2DEG and the SAW becomes quite weak.

\section{discussions}
We have study the interaction of SAW with the 2DEG by introducing
the piezoelectric coupling. It is found the the coupling is most
strong when the SAW transport along the direction of $\theta\sim
25^0$ while it reaches a minimum when transport along the
$\theta=45^0$ direction. This result is very helpful for
experimentalist when designing devices for different purposes of
SAW transport on piezoelectric crystals.

\section{acknowledgement}
This work was supported in part by NSF of China. S.-J. Yang was
supported by a grant from Beijing Normal University.

\centerline {Figure Captions}

Figure 1 A Schematic description of the sample on which the SAW
transports. The 2DEG is located a distance $d$ beneath the
surface.

Figure 2 The velocity of the SAW versus the transport direction
without piezoelectric coupling.

Figure 3 Piezoelectric coupling coefficient of the SAW with 2DEG
under various transport directions. As the transport angle
increases, the coupling constant experiences first a rise and then
a fall process. At $\theta=45^0$, the coupling becomes quite weak.

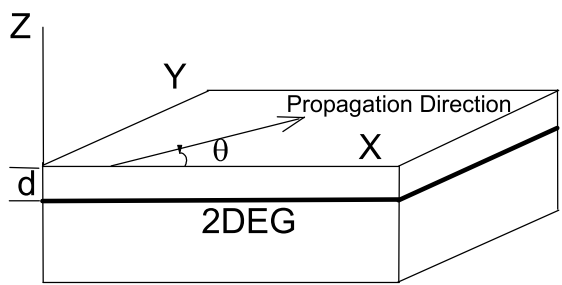
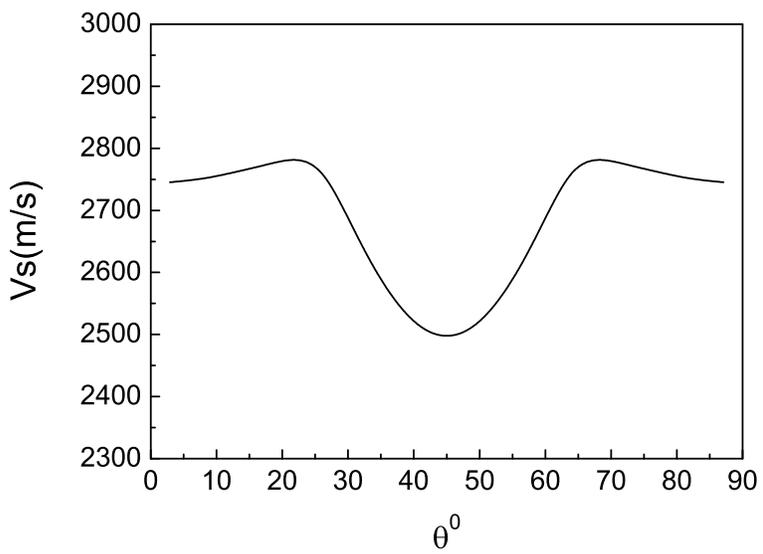
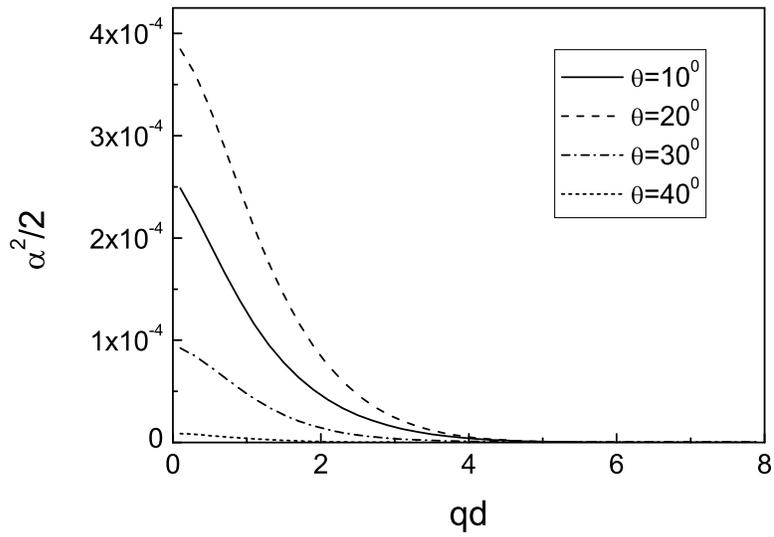
\end{document}